# Recent Advances of Image Steganography with Generative Adversarial Networks

Jia Liu[*], Yan Ke, Yu Lei, Zhuo Zhang, Jun Li, Peng Luo, Minqing Zhang and Xiaoyuan Yang

*liujia1022@gmail.com*

**Abstract**: In the past few years, the Generative Adversarial Network (GAN) which proposed in 2014 has achieved great success. GAN has achieved many research results in the field of computer vision and natural language processing. Image steganography is dedicated to hiding secret messages in digital images, and has achieved the purpose of covert communication. Recently, research on image steganography has demonstrated great potential for using GAN and neural networks. In this paper we review different strategies for steganography such as cover modification, cover selection and cover synthesis by GANs, and discuss the characteristics of these methods as well as evaluation metrics and provide some possible future research directions in image steganography.

## 1 Introduction

Steganography hides secret messages in a file. The term steganography is also known as secret writing.[1] In cryptography, the obvious visible encrypted information, no matter how unbreakable, will attract the attention of attackers and may be subject to more control in an encrypted environment. Compared to cryptography, steganography has the advantage that the existence of secret information itself will not be suspected. [2]. Media with secret messages is called stego media, and the media used to hide messages but does not contain secret messages is called cover media. For attackers, they use steganalysis techniques in the hope of preventing the transmission of secret information.

In modern steganography [1] , according to the different construction strategies of the stego media, Steganography techniques are divided into the following three types. *Cover modification* steganography achieved the purpose of hiding messages by making use of redundant information in the digital carrier and modifying this redundant information. However, the modification will inevitably lead to the difference in the distribution between the original cover and the stego carrier. *A cover selection* steganography which is somewhat similar to image retrieval, which is to implement steganography by selecting a natural, unmodified, normal carrier in a large database that can extract messages as a stego medium. This method has a very low payload so that cannot be applied to practical applications. *Cover synthesis* steganography refers to the construction or synthesis of a seemingly normal media containing secret messages to achieve information hiding. However, about ten years ago, constructing realistic digital cover media is more of a theory, and there is no practical technical solution. This status quo restricts the development of cover synthesis steganography.

Fortunately, a powerful generation model, generative adversarial network, was proposed in 2014 [3] . The generator in the GAN model is capable of synthesizing realistic images. There are two main directions for the study of GAN. One direction is to optimize the model of [4-7] from different aspects such as information theory [8] and energy-based model [9]. The other research direction is to try to apply GAN to more research fields, such as computer vision (CV) [4] and natural language processing (NLP)[10]. [11-13] reviews recent GAN models introduces applications. However, But these review articles do not focus on a specific application. In this article, we focus on GAN's research progress in the special field of image steganography.

Recently, there are many articles using GAN to design steganographic methods [14-19], which enrich the technical means of steganography. The main purpose of this article is to try to discuss the role of GAN in image steganography, and point out the problems faced by GAN-based image steganography. As we will see, current steganography methods using GAN have covered three strategies of current traditional steganography techniques, i.e. modification methods, selection methods, and synthesis methods. In this paper, we start from the basic model of steganography and briefly review the traditional steganography methods and security issues. Then, after introducing the basic concept of GANs, we give a detailed discussion of three strategies in image steganography with GAN. Besides these GAN-based methods, there are some ways to design steganography schemes using deep neural networks [20] or adversarial samples [21], which will be mentioned briefly. To the best of our knowledge, this is the first article that attempts to address the application of GAN in the field of image steganography. We try to explain the role of GAN in image steganography according to the classification of traditional steganography. The most important result of using the deep generative model in image steganography is that, the traditional steganography design pattern that relies on the manual method will be transformed into a pattern that relies on automatic data-driven steganography scheme. On the basis of the generative model, cover synthesis and cover selection steganography have some more attractive properties. We also discuss the possible reasons why a powerful generative model can perform promising in information hiding tasks and its role in our goal to steganography.

The rest of this paper is organized as follows, the classical steganography model as well as strategies and security criteria are introduced in Section 2. We also give a brief overview of the implementation of traditional steganography schemes, focusing on the characteristics and performance of these methods. In Section 3, we briefly review the basic ideas and applications of GAN. Then In Section 4, we discuss several methods in cover modification strategy using GAN and some research directions for improvements. In Section 5, we discuss a special cover selection strategy using GAN. In Section 6, we then give a detailed discussion in cover synthesis steganography with GAN. Some possible research directions are discussed. In Section 7, we will give some evaluation metrics for images steganography by GAN. A short conclusions and perspective are given in Section 8.

**2 Steganography Preliminaries**

**2.1 Steganography Model**

The classical steganographic model is the prisoner's problem [22] with three participants, as illustrated in Fig.1. Both Alice and Bob are held in separate cells. They are allowed to communicate with each other, but all their communications are monitored by the warden Wendy. In modern steganography, ever channel between Alice and Bob contains five elements: cover source *c*; data embedding/extraction algorithm ***Emb/Ext***, secret key ***k*** for embedding/extraction, message ***m***, and communication channel, as shown in Fig.1.

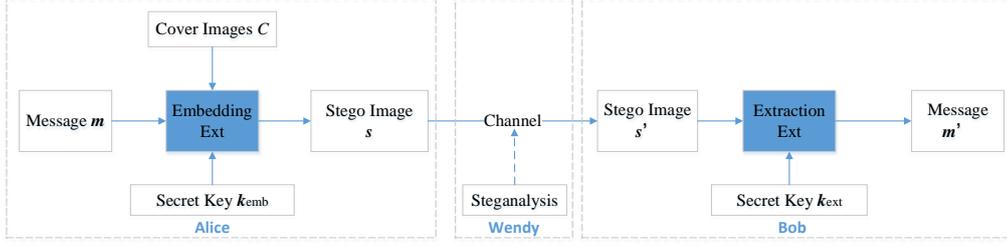

Fig. 1. The prisoner's problem model for steganography

Using a data embedding method Emb(·), based on a specific carrier $c$ or a set of carriers $C$, Alice needs to design a scheme to construct a stego media $s$ with an embedding key $k_{emb}$. The generation process of the stego carrier $s$ can be expressed as:

$$s = \text{Emb}(c\,|\,C, m, k_{emb}) \tag{1}$$

For Bob, the stego media he receives can be expressed as $s'$. He can also recover a secret message $m'$ using an extraction key $k_{ext}$ and message extraction operation Ext(·).

$$m' = \text{Ext}(s', k_{ext}) \tag{2}$$

The message extraction key and the embedded key can be different with public key steganography [23]. In this paper, we only focus on the symmetric steganographic algorithm, where $k_{emb} = k_{ext}$ is assumed. When $s'=s$ is guaranteed, the steganographic channel is lossless. The above Eq. (1) and Eq. (2) only describe the process of message embedding and extraction. For the steganographic task, the core requirement is that the constructed stego media $s$ must be indistinguishable from the cover image $c$ or the cover set $C$ to realize the task of message hiding. Here we define an abstract distance metric $D_{distinguishable}$ to represent indistinguishability:

$$D_{Distinguishable}(C_{cover}, S_{stego}) \leq \varepsilon \tag{3}$$

where $C_{cover}$ and $S_{stego}$ represent the cover set and the stego set respectively, $\varepsilon$ represents a quantifiable level of security for indistinguishability, ε-security. The above three expressions indicate the basic goal of a steganographic algorithm, which we called them steganographic basic conditions (SBC).

In order to facilitate the transmission of secret information, the embedded capacity of the steganographic system [24] should be high enough. There are already many evaluation criteria for measuring message capacity such as per pixel bits, or the ratio of secret messages to cover media, and so on.

**2.2 Steganography Security**

Steganography security depends on the means of the attacker, According to Wendy's work in examining the media, she can be active or passive. If Wendy only checks whether the stego media is natural or normal in the channel transmission, she is called an active warden. If Wendy tries to detect the existence of covert communication by extracting secret messages directly, she is a passive warden. Many reviews of steganography focus solely on the passive warden mode. In practice, it is common for Wendy to have both active and passive responsibilities as a warden. According to the Kerchhoffs's principle[25] of security systems, Wendy should have all the information about the steganography algorithm, which means she knows every detail of how the carrier object is used by both sides. For steganographer, the security of steganography-system mainly includes two aspects.

*Active attack*: In the case of an active warden, steganographic security is mainly concerned with the difficulty of message extraction. The traditional realization of steganography that lacks shared secrets is through obscure security forms. Hopper [26] and Katzenbeisser[27] independently proposed the complexity theory definition of steganographic security. In our recent work[28], a stego-security classification is proposed based on the four levels of steganalysis attacks:

a) Stego-Cover Only Attack (SCOA): In this case, we assume that the steganalysis attacker can only access a set of stego-covers.

b) Known Cover Attack (KCA): In this case, being able to perform SCOA, the attacker can also obtain some original cover carriers and their corresponding stego carriers. Within polynomial complexity, the number of pairs is limited.

c) Chosen Cover Attack (CCA): In this situation, an attacker can use the steganographic algorithm to perform multiple message embedding and extraction operations with a priori knowledge under KCA. Within polynomial complexity, the number of invocation operations is limited.

d) Adaptive Chosen Cover Attack (ACCA): The ACCA mode means that when the CCA mode challenge fails, another CCA attacks can be performed until the attack is successful.

Under this definition, the steganalyzer does not need to know the probability distribution of the cover, but only assumes that Wendy can access to a black box to generate the cover. She can sample the cover from the black box. Meanwhile, steganographic security is established through the adversarial game between warden and judges. This method is based on the classification standard of security level in cryptography. However, the difficulty of constructing this black box limits the development of security based on computational complexity in the case of active attacks. Fortunately, as we will see, the generative model provides a technical basis for constructing this black box, and security evaluation criteria based on complexity theory will play a greater role in the evaluation of steganographic security.

*Passive attack:* In the case of a passive attack, the indistinguishability between steganographic and natural vectors is the key issue to steganography security. As the indistinguishability includes the imperceptibility for the human visual system and the undetectability for machine statistic analysis system. Therefore, we have

$$D_{\text{distinguishable}}(C_{\text{Cover}}, S_{\text{Stego}}) = D_{visual}(C_{\text{Cover}}, S_{\text{Stego}}) + D_{\text{statistical}}(p_{\text{cover}}, p_{\text{stego}}) \quad (4)$$

where $D_{visual}(C_{Cover}, S_{Stego})$ denotes the perceptibility by human and $D_{statistical}(p_{cover}, p_{stego})$ denotes the statistical distance between distribution of cover images and distribution of stego images. In terms of human vision, most current steganography method can achieve indistinguishable between stego medium and normal medium, which can be represented as $D_{visual}(C_{\text{cover}}, S_{\text{stego}}) = 0$. Statistical indistinguishability is the most studied area of steganographic security. Cachin[29] defined a quantified security for a steganography scheme by the relative entropy between the cover distribution $p_{cover}$, and stego distribution $p_{stego}$:

$$D_{\text{statistical}}(p_{\text{cover}}, p_{\text{stego}}) = D_{KL}(p_{\text{cover}} \parallel p_{\text{stego}}) = E_{p_{\text{cover}}}[\log \frac{p_{\text{cover}}}{p_{\text{stego}}}] \quad (5)$$

Based on this definition, if $D_{KL}(p_{cover}\|p_{stego}) \leq \varepsilon$, steganography system is called ε-security. If $\varepsilon = 0$, the scheme is called perfectly secure. Although the definition of security based on information theory is popular, it is an ideal way to define security regardless of its implementation. It requires the assumption that Wendy fully understands the probability distribution of the cover and stego sets.

At the same time, there are other ways to define steganographic security. In ref. [30], ROC performance is adopted as an alternative security measure, the steganographic security is defined with the practical performance of the steganalysis. A statistical method, Maximum mean discrepancy (MMD)[31], for testing whether two classes of samples are generated from the same distribution, is also be considered as a measure of steganographic security, The advantage of the MMD method is numerically stable even in high-dimensional space.

**2.3 Strategy Implementation**

In this paper, we assume that the embedding algorithm associates every message *m* with a pair [*s*, π], where *s* is stego image that can be obtained ( by cover modification, selection or synthesis) from a set of all stego images *S*, the π is the probability distribution for a specific embedding operations, π(*s*)=P(*S* = *s*|*m*). Unlike the [32], in this paper we do not give the original cover *c* explicitly, we treat image steganography as a mapping process from message *m* to stego image *s*.

If Bob receive *s*, Alice could send up to

$$H(\pi) = \sum_{s \in S} \pi(s) \log \pi(s) \tag{6}$$

bits message on average. In this situation , the average distinguishability can be also denoted by :

$$E_\pi[D_{distinguishability}] = \sum_{s \in S} \pi(s) D_{distinguishability}(s) \tag{7}$$

where $D_{distinguishability}(s)$ is a metric indicating that the cover image *s* is indistinguishable from the natural cover image *c*. Similar to [32], the task of embedding can assume two forms:

**Distinguishability-limited sender**: In this mode, the average payload will be maximized given a fixed indistinguishability:

$$\arg\max_\pi H(\pi) \quad \text{st.} \ E_\pi[D_{distinguishability}] \tag{8}$$

**Payload-limited sender** : In this mode, the indistinguishability metric is minimized given the size of the transmitted message.

$$\min_\pi E_\pi[D_{distinguishability}] \quad \text{st.} \ H(\pi) = m \tag{9}$$

The **Payload-limited sender** is commonly used in cover modification steganography in which $D_{distinguishability}$ is often replaced by a defined distortion function. This is due to that there is a lot of redundancy in image cover, minimizing $D_{distinguishability}$ indicate modification operation introduce least abnormities in stego image. In fact, distinguishability-limited sender is more in line with the original intention of steganography. This is because the redundancy present in the image can carry different message capacities. We will see that, in cover synthesis steganography scheme, if we can find a procedure to create stego image with a fixed average distinguishability, maximize the average payload will be the core aim. In the next section, we will give detail of steganography scheme.

When designing a practical steganographic scheme, it can be divided into three different fundamental architectures according to the different ways of obtaining the stego image, they are cover modification, cover selection, and cover synthesis.

**2.3.1 Cover Modification**

There exist two mainstream approaches for cover modification steganography: One approach is to maintain the invariance of a statistical model [31] and the other type of methods implement embedding by minimizing a certain distortion function. [32], as shown in Fig. 2.

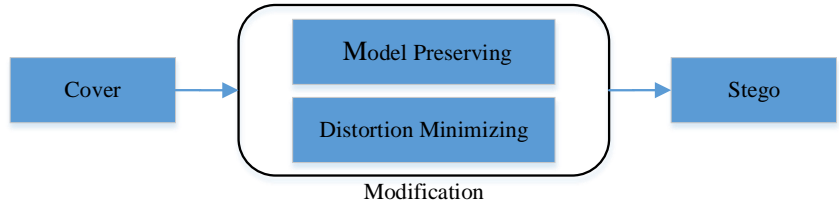

Fig. 2 .Steganography with cover modification

A steganographic strategy that maintains a certain statistical model is not safe enough in the face of well-designed steganographic features [33, 34]. Steganography based on minimizing distortion is more straightforward and attractive. It abandoned the need for statistical modeling of the cover source and instead sought to reduce the distortion [32, 35, 36] introduced by the embedding. The method based on minimizing distortion is state-of-art in steganography with cover modification. This method has a high embedding capacity and is simple and convenient to implement. The distortion function is usually a simple additive distortion. Some improved distortion functions are also proposed [37-39].

However, the definition of distortion is too vague to accurately define the impact caused by a modification. Furthermore, although stego $s$ is highly correlated with specific cover $c$, a well-trained classifier that training on data set $C_{cover}$ and $S_{stego}$ are able to perform steganalysis. Methods of carrier modification always assume that modification can avoid the attention of human vision, $D_{visual}$ ($C_{cover}$, $S_{stego}$) ≈ 0. Modification inevitably leads to the difference between the carrier distribution and the stego carrier distribution. $D_{statistical}$ ($p_{cover}$, $p_{stego}$) ≠ 0 .In the case of passive attacks, the relationship between distortion $D_{distortion}$ and statistical distinguishability $D_{statistical}$ is far from clear.

**2.3.2 Cover Selection**

Cover selection methods can be divided into two ways. One is to select a candidate image for modification [40-42], these methods look for a suitable cover in the database to implement cover modification steganography. Although these methods are called cover selection, they are still essentially a cover modification method, we do not treat these schemes as cover selection steganography. The other is to select a cover image as a stego image without modification as shown in Fig. 3.

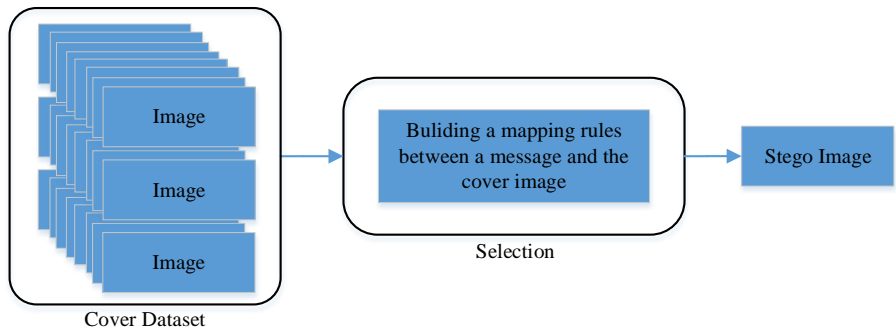

Fig. 3. Steganography with cover selection

The essence of the cover selection method is to establish the mapping rules between a message and a stego image. Zhou et al. [43] introduce a cover selection steganography scheme by using the bag of words model [44] (BOW) in computer vision. To hide information into an image, visual words are extracted to represent the secret message. Visual words from an image set are extracted using a BOW model, and a mapping relation between keywords in the message and visual words in the image is established. According to the known message and a set of rules, the selection method looks for the image that can extract the message as a cover image in the image dataset. This set of rules is essentially a secret key for cover selection steganography. However, as the mapping relationship between message and stego is fixed and the mapping structure is usually quite simple, it is easier to deduce the mapping rules between message and stego through some observations under the active attack. Another problem is that this simple mapping rule leads to extremely low embedding rates, which hinders the deployment of such algorithms in practical applications.

**2.3.3 Cover Synthesis**

The third strategy is based on image synthesis. In this method, Alice tries to create a new image to carry the required secret information. If we can make sure that synthesis image is real enough that $D_{indistinguishable}$ ($C_{cover}$, $S_{stego}$) = 0, then we can achieve a secure steganographic system theoretically. At present, there are two kinds of steganographic methods with image synthesis, as shown in Fig. 4.

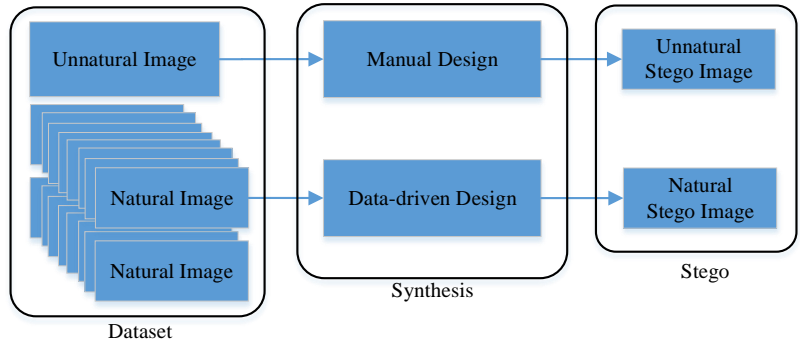

Fig. 4. Steganography with cover synthesis

Since the realistic image synthesis is a difficult problem until now, traditional cover synthesis method tried to achieve steganography task via unnatural image synthesis, texture image[45] and fingerprint image[46]. Otori and Kuriyama [47, 48] first try to combining information hiding with pixel-based texture synthesis. Wu et al.[49] proposed a reversible texture image synthesis for steganography. Qian et al.[50] propose a steganography method which secret messages are hidden in a texture image during the process of synthesizing. [51, 52] introduce a deformation-based texture for information hiding. Zhang et al. [53] propose a construction-based steganography scheme which conceals a secret message into a fingerprint image. The premise of the steganography based on texture synthesis is to assume that the stego carrier can be an image without semantic information, which limits the application of texture synthesis steganography in a larger field.

The other approach is to train a generator by a generative model with a large amount of data. Stego images can be obtained from the realistic image generator. A probability distribution which described by the generative model is $p_{model}$ or $p_g$. In some cases, the model estimates $p_{model}$ explicitly. Furthermore, if the images obtained by the generative model are treated as the stego images, the distribution of the generated samples can also be denoted by $p_{stego}$. Maximum likelihood estimation is used for estimating parameters of density function that is computationally tractable, while variational methods and sampling methods such as Markov chain Monte Carlo are used for density function that is intractable, requiring the use of approximations to maximize the likelihood. Because of the complexity and high dimension of natural images, it is impossible using an explicit density function to describe the distribution of natural images. Fortunately, the GAN model uses an indirect method to obtain the distribution of real images, which does this by generating samples rather than estimating the specific form of the distribution. In Section 6, we will see how to design the steganography method using an image generator which is obtained by training GAN model.

**2.4 Summary on Traditional Steganography**

Under passive attack, perfect steganography method aims to find an algorithm that satisfying steganography condition $p_{stego}= p_{cover}$. From the point of view of technical feasibility for steganographic security, Fridrich[1] ignores the fundamental question of whether it is feasible to assume that the distributions $p_{cover}$ and $p_{stego}$ can be estimated in practice or even whether they are appropriate descriptions of the cover-image source. Therefore, the choice of the specific form of $p_{cover}$ actually becomes the key issue in designing of steganography method. Traditional steganography research has focused on methods based on cover modification.

Most of the methods based on modification try to ensure that the modification operation should keep the invariance of a specific statistical characteristic, that is: $p_{cover} \neq p_{cover\_specific}$, $p_{cover\_specific} = p_{stego\_specific}$, The disadvantage is that the opponent can usually identify the statistic beyond the selected model fairly easily, which allows reliable detection of embedded changes.

In the steganographic scheme of cover synthesis, the distribution of the stego images $p_{stego}$ should be close enough to the real distribution of the natural cover image $p_{real}$. But the real image of distribution can hardly be given explicitly. We can only approximate the real image distribution $p_{real}$ by describing the distribution of existing data, $p_{real} \approx p_{data}$. As discussed in the previous section, GAN allows us to train a generator with an adversarial learning model. The distribution of the samples sampled from the generator satisfies $p_g=p_{data}$. When we get the stego image directly from the generator, we can achieve statistical indistinguishability. When we can get a proper description of the cover image source, we hope that the steganography algorithm meets not only the indistinguishability security but also Kerckhoffs's principle. Although this paper will mainly consider steganography under passive attack, we will also introduce steganography under active attack in some methods. In order to understand the characteristics of GAN-based steganography methods clearly, the following Section 2 will briefly discuss the basic principles and characteristics of GAN. As we will see, the basic fundamental questions which are neglected by traditional steganography is exactly what GAN wants to solve.

**3 GAN Preliminaries**

**3.1 Core concepts**

The basic idea of GANs is an adversarial game between two players, as illustrated in Fig. 5. The task of generator $G$ is to transform the input noise $z$ into a sample $G(z)$. The discriminator $D$ determines whether the generated fake sample is indistinguishable from the real sample. A generative model $G$ is actually a neural network with parameters $\theta$ denoted as $G(\mathbf{z}; \theta)$. The output of the generator can be thought of as sampling from a distribution $G(\mathbf{z}; \theta) \sim p_g$. With a lot of real data $x$ drawn from $p_{data}$, the goal of generator training is to make the generator's distribution $p_g$ close to the real data distribution $p_{data}$.

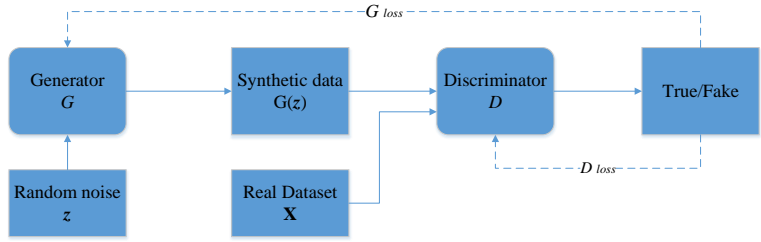

Fig. 5. The general structure of GAN.

Goodfellow et al use a multilayer perceptron as a generator. The objective function is shown in Eq. 10:

$$\min \max V(D,G) = E_{x \sim p_{data}(x)}[\log D(x)] + E_{z \sim p_z(z)}[\log(1-D(G(z)))] \tag{10}$$

They also show that the optimization process can be seen as minimizing Jensen-Shannon divergence (JSD) [3] between real data distribution and generator distribution. More importantly, if both generator and discriminator have adequate capability, the game will converges to its equilibrium with $p_g = p_{data}$. In practice, the parameters for two networks are updated in the parameter space.

### 3.2 Improvements and Application

#### 3.2.1 Improvements

The improvements of GAN models can be classified into two aspects: the architecture and the loss function. To be specific, GANs are classified into different types as shown in Fig. 6.

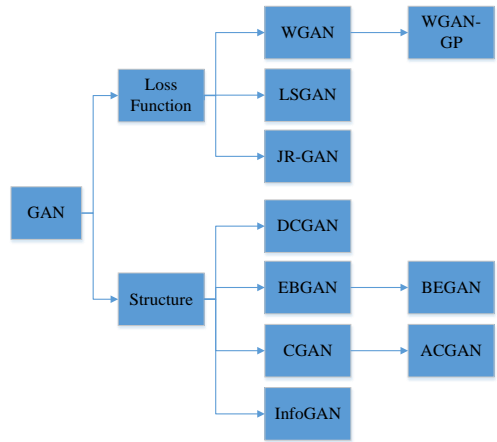

Fig. 6. Improvements on GAN models

The most famous model is DCGAN. [4] which performs well in image synthesis in the early work of the research. In order to control generated result, different GAN such as CGAN[54], InfoGAN[8], ACGAN[55] are proposed. Some methods have been proposed for solving the model collapse problem by designing a new loss function such as mini-patch feature [5], MRGAN [56], WGAN [6] and WGAN-GP [57].

#### 3.2.2 Applications

The application of early GAN was mainly concentrated in the field of computer vision, such as image inpainting [58], captioning [59, 60], detection [61] and segmentation[62]. GAN also has some applications in the field of natural language processing, such as text modeling [10, 63], dialogue generation [64] and machine translation [65].

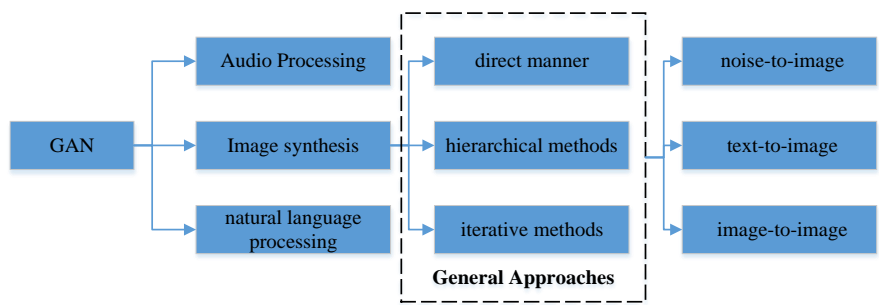

Fig. 7. Applications with GAN models

Huang et al. [13] summarize main approaches in image synthesis into three methods, i.e. direct methods, hierarchical methods and iterative methods. Direct Methods such as GAN, DCGAN, Improved-GAN [5], InfoGAN, f-GAN [66] and GANINT-CLS [67], usually using one generator and one discriminator. The hierarchical approach uses two generators and two discriminators. The idea behind this approach is to divide the image into two different pieces of content, such as "style and structure" and "foreground and background." Hierarchical methods refers to the model which generates images from coarse to fine using multiple generators with similar or identical structures.

On the other hand, depending on the source of the generated image, image synthesis can also be divided into three different synthesis methods, namely noise-to-image, text-to-image and image-to-image. Text-to-image synthesis is a research field with great prospects. It means that machines can understand the semantic information of text. GAN provides us with a promising text-to-image synthesis method, such as GAN-INT-CLS [67], GAWWN [68], StackGAN [69] and PPGN [70]. The GAN-based approach so far produces images that are sharper than any other generation method. Image translation is related to style transfer [71], which constructs a generated image with specific content and style by using a content image and a style image. Image-to-image translation by GANs has also been successfully applied in some image or video generation applications[72].

**3.3 Steganography by GAN**

In this section, we first summarize the characteristics of GAN. In fact, GAN's characteristics can be viewed from the following three aspects: an adversarial game, a generator or a mapping function. These consistent with the classification of steganographic basic strategies, i.e. cover modification, cover synthesis and cover selection. Then, we summarize the three main types of approaches mentioned in this paper based on the above characteristics as shown in Fig. 8.

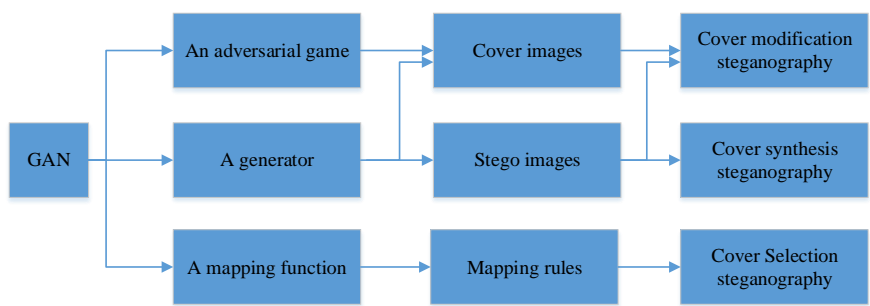

Fig. 8. The categories for steganography from the point of view of GAN

Under the first view, GAN is treated as an adversarial game between generator and classifier, both generator and discriminator are equally important. This kind of viewpoint pays attention to the whole process of the adversarial game and pays more attention to the positive effect produced by the discriminator. In fact, there have been some studies on steganography methods based on game theory before the GAN is proposed, such as [73-75]. The family by modification-based steganography take advantage of the concept of a game simulation between two-players: Alice-agent and Eve-agent. Historically MOD [76] and ASO [77] were the algorithms of this type. Recently some researchers take advantage of the adversarial concept by generating a fooling example (see for adversarial example[78]), but those approaches are not an adversarial game between generator and discriminator by GAN. However, unlike GAN using iterative and dynamic game process, those approaches are not a dynamic process, there is no dynamic adversarial game simulation. They are not trying to reach a Nash equilibrium, and only considered the implications of steganalysis at the beginning of designing steganographic schemes. On the other hand, this traditional game strategy is more of theoretical analysis, while the steganography based on GAN can be used in designing practical steganography scheme. The game between steganalysis and steganalysis is similar to the game between generator and discriminator in GAN model. Inspired by this similarity, GAN is chosen to improve the performance of the traditional cover modification steganography method. We will discuss the specific method of this part in detail in Section4.

The second view treats the generator training procedure in GAN as a powerful construction method of the mapping function. This mapping function maps a driving signal through a neural network to an image which belongs to a specific image set. For image steganography, the ability to construct the mapping function enables GAN to be applied to the cover selection steganography scheme for constructing the mapping between message and cover, which is an interesting idea. We will elaborate on the details of this scheme in Section5.

The third view is to regard GAN as a method to construct a powerful generator. As we all know, this view treats the result of the game process, a powerful generator, as the most successful innovation of GAN model. A much more interesting approach using cover synthesis is to directly generate images that will be considered stego images. This kind of method mainly takes image synthesis as basic tools. In fact, image steganography based on cover synthesis is a special application of image synthesis. The key issue raises how to hide the message in the synthetic image. A typical approach is to obtain a stego image by introducing steganography constraints or a loss term with message extraction. This is the basic idea of the steganography based on image (or cover) synthesis. We will discuss some recent researches in detail in Section 6.

**4 Cover Modification by GAN**

As mentioned above, cover modification by GAN focuses on the adversarial game between steganography and steganalysis. These methods use generator trained by GAN to construct various core elements in the cover modification scheme. One strategy is to generate the original cover image, the second is to generate the modification probability matrix in the framework of minimizing distortion, and the third is to directly use the adversarial game among tripartite, such as Alice Bob and Wendy, to learn a modification algorithm for generating stego images.

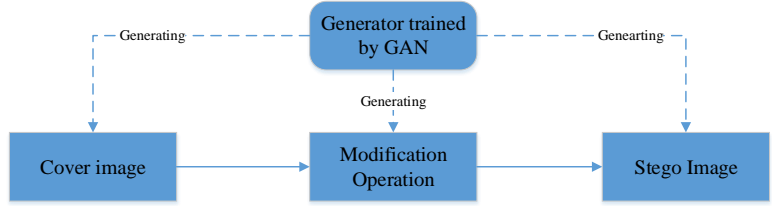

Fig 9. The categories for cover modification by GAN

### 4.1 Generating Cover Images

Volkhonskiy et al.[79] proposed the application of GAN to steganography. They construct a special generator for creating cover-image, synthetic images generated by this generator are less susceptible to steganalysis compared to covers. This approach allows generating more steganalysis-secure cover that can carry message using standard steganography algorithms such as LSB or STC. They introduce the Steganographic Generative Adversarial Networks which called SGAN consists of three networks. A generator $G$ generate, a discriminator $D$ and a steganalysis classifier $S$ determines if an realistic image hiding a secret information. The workflow of SGAN is illustrated in Fig.10.

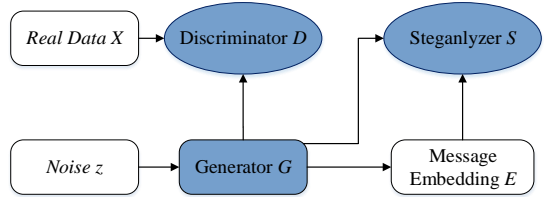

Fig. 10. SGAN workflow diagram

SGAN trains $G$ with $D$ and $S$ simultaneously. We can get the game as follow:

$$L = \alpha \left( E_{x \sim p_d(x)}[\log D(x)] + E_{z \sim p_z(z)}[\log(1 - D(G(z)))] \right) +$$
$$+ (1 - \alpha) E_{z \sim p_z(z)}[\log S \left( \text{Stego}(G(z)) \right) + \log(1 - S_D(G(z)))] \to \min_G \max_D \max_{S_D} \quad (11)$$

Where parameter $\alpha$ [0; 1] denotes the weight between the quality of the generated image against the steganalysis, $S(x)$ is the probability for $x$ is stego image.

Similar to SGAN, Shi et al. [17] use the same strategy that generates cover images for steganography with adversarial learning scheme, named SSGAN. The SSGAN architecture also has one generative network called $G$, and two discriminative networks called $D$ and $S$. Compared with the SGAN, WGAN is introduced for generating images for higher quality and improving training process. A more complex network called GNCNN[80] is chosen as the discriminator $D$ and the steganalyser $S$.

Another interesting cover image generation method is proposed by Wang et al.[81] as shown in Fig. 11. Unlike SGAN and SSGAN, a discriminator $D$ determines whether an image is stego or real image, stego($G(z)$) and real image sample $x$ are used as the input of discriminator model $D$. Such improvement makes the distribution of the stego image closer to the real data distribution. An interesting result of this scheme is that the images generated directly by the generator may not be realistic, and the fidelity of the stego image is achieved after the modification operation.

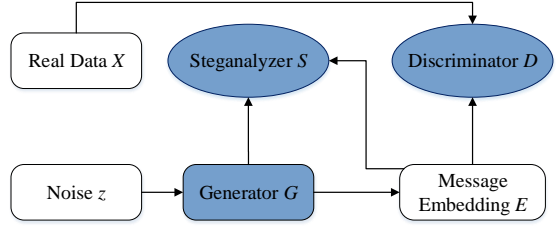

Fig. 11 Stego-GAN workflow diagram

### 4.2 Learning Distortions

Tang et al. [82] Tang et al. proposed to ASDL-GAN model to automatically learn a distortion function. This scheme follows the state-of-art steganography by minimizing an additive distortion function[32]. The change probabilities matrix $P$ can be obtained via minimizing the expectation of the distortion function [83]. The generator $G$ in their scheme is trained to learn the change probabilities $\mathbf{P}$ for input image.

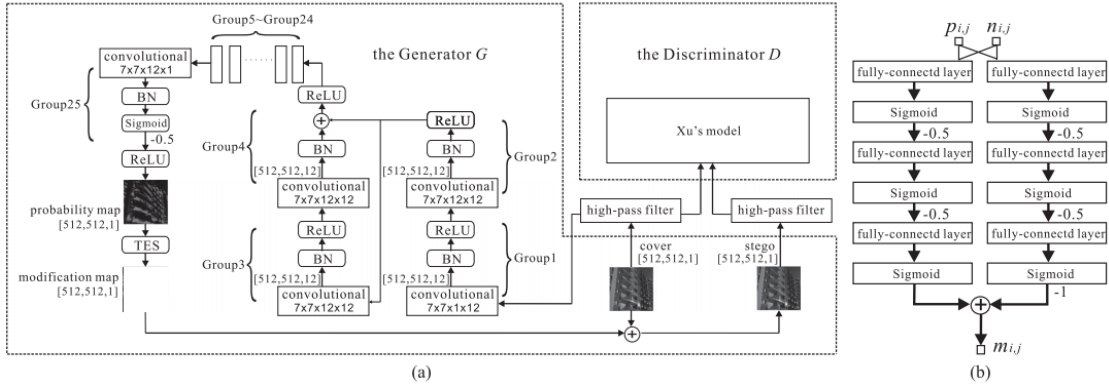

Fig.12 (a) Architecture of the ASDL-GAN framework[82].(b) The structure for TES activation function.

As illustrated in Fig. 12(a), , the discriminator $D$ in ASDL-GAN framework adopts the Xu's model architecture [84]. The embedding simulator (TES), is used as the activation function in the training procedure. The reported experimental results showed that steganographic distortions can be learnt by ASDL-GAN.

Inspired by ASDL-GAN, UT-SCA-GAN[14] proposed by Yang et al. with same component modules as ASDL-GAN: a generator, an embedding simulator, and a discriminator. Compared with the ASDL-GAN, Tanh-simulator, an activation function, is used for propagating gradient. In addition, a more compact generator based on U-NET [85] has been proposed. The experimental results show that this framework can improve the safety performance. At present, there is no guarantee [86] that the probability map obtained will defeat the security performance of HILL or S-UNIWARD with STC in practice. It is also unclear whether the loss of the generator must incorporate terms related to safety and terms of payload size.

### 4.3 Embedding as Adversarial Samples

Some researchers have also designed steganography with the idea of adversarial examples [87]. However, simply adding perturbations directly to a stego images can also result in instability of message extraction. Tang et al. [88] proposed a steganography scheme called adversarial embedding (ADV-EMB), which tries to modify the original cover for message hiding while fooling a steganalysis classifier. The ADV-EMB scheme is illustrated in Fig. 13.

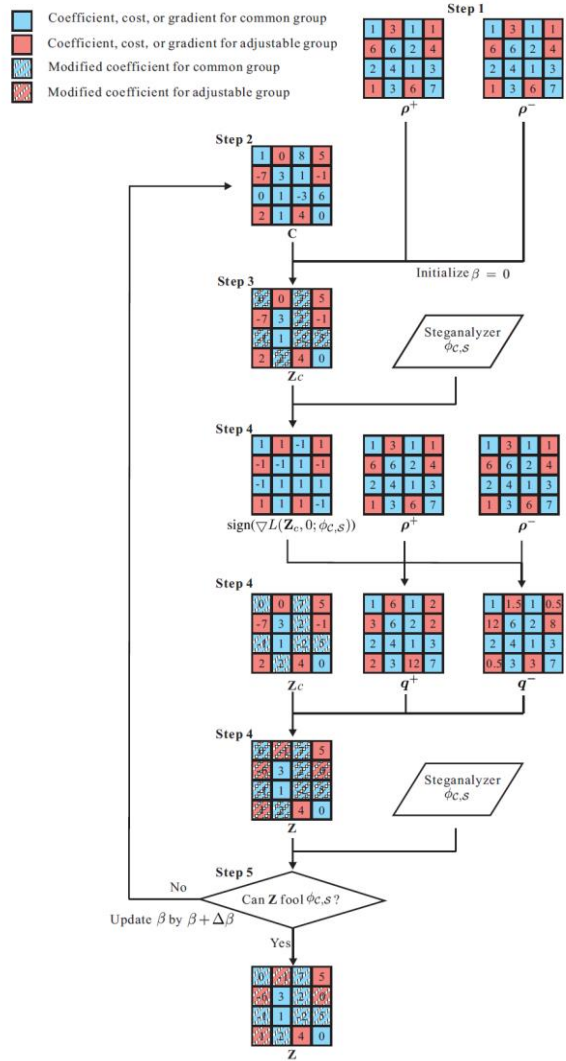

Fig. 13. The model architecture of ADV EMB scheme [88].

The pixels of candidate stego image are divided into two groups, one group of pixels is used for modification-based embedding, and a tunable group of pixels is used for perturbation as an adversarial sample to resist steganalysis. ADV-EMB adjusts the cost of modification operation with back-propagation on the gradient of the steganographic analyzer. Their experiments show that the ADV-EMB achieve better security performance.

Similar to [88], Ma et al. [89] modify the image pixel following the adversarial gradient map while embedding. The adversarial gradient map is the matrix generated from the neural network model and has the same size as the cover image. Each element of the adversarial gradient map is the gradient value that make the steganalyzer tend to have a false classifying result.

**4.4 Summary on Cover Modification**

SSGAN [17]construct a special cover-image generator, they can use standard steganography algorithms such as LSB or for information hiding. [82]and [14] train a generator of modification probabilities matrix for minimizing a suitably defined additive distortion function. [15, 78, 88, 89] learning a whole cover modification steganographic algorithm using GAN. They focus on the *adversarial game* between steganography and steganalysis. They both introduce a steganalyzer against the steganography either explicitly or implicitly. Although these methods have even achieved better anti-analysis capability than traditional steganography methods. These methods are still faced with traditional security threats when Wendy can get all the information of the algorithm to obtain stego and cover. In theory, these methods can't resist more powerful steganalysis tools, since the embedded operation will inevitably cause some abnormal features

## 5 Cover Selection by GAN

The essence of the cover selection method is to establish the mapping relationship between message and cover. As far as we know, there is very little literature on this subject that attempts to use GAN to design the steganography scheme, Ke et al. [16]made a preliminary attempt on this subject.

### 5.1 Generative Steganography by cover selection

Ke et al.[16]propose a Generative Steganography method which meets Kerckhoffs' principle (GSK), The idea is that the sender establishes a mapping relationship using generator between the message and the selected image. For the receiver, message is directly generated by the selected image. In [28], this method is also called cover first generative steganography (CFGS). The essence of this method is to establish a mapping relationship between cover image and secret message so that a cover image will naturally turn to be a stego image. Statistical steganographic analysis does not work because there is no operation for cover modification. Ke et al. establish a mapping relation between message and cover image using GAN. To ensure the security, Kerckhoffs' principle is also introduced in their **GSK** method. Fig. 14 shows the three message extraction scenarios under this framework.

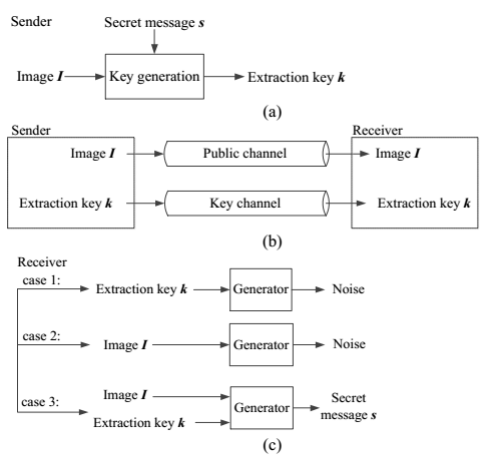

Fig. 14. The model architecture of GSK method [16].

As for the receivers (Fig.14c): *Case 1*, the only *k* is received corresponding to a failed message extraction, only noise can be recovered. *Case 2*, only *I* is received corresponding to an intercept from attackers, there is also noise output for attacker. *Case 3*, when *I* and *k* are both obtained, the message *s* could be recovered. In order to realize that aim, two mapping relationship between the key *k* and the message *s*, and relationship between the cover *I* and the message *s* should be constructed by a Message-GAN and Cover-GAN, respectively. Message-GAN which implement by InfoGAN [8] is to use feature codes to control the output. Cover-GAN which is similar to Abadi [90]method for cryptology is used to determine the generation of the message $s$.

**5.2 Summary on Cover Selection**

There are few studies on cover selection steganography based on generative models. This types of approach treats the generator as a mapping between a message and an existing natural image. The advantage of this method is that the image is 100% natural due to no modification. For the moment, the low embedding capacity of cover selection steganography is still a bottleneck to its development.

**6 Cover Synthesis by GAN**

The method of generating cover images by generator trained by GAN is essentially a kind of image synthesis. In our opinion, the key of steganography by image synthesis is that the stego image should be obtained directly from a black box, such as a generator. Since the biggest advantage of GAN is the ability to generate realistic natural images, we will see in this section how to use generators to generate stego images. In this case, message extraction will be referred to as the most critical issue in image synthesis steganography.

**6.1 Supervised stego image synthesis**

Similar to Abadi [90], Hayes et al. [15] try to use a neural network to learn a steganography algorithm with adversarial training. In their framework, the three players, Alice, Bob and Eve, are neural networks. $\theta_A$, $\theta_B$, $\theta_C$ denote the parameters for the networks, respectively. The full scheme is depicted in Fig.15.

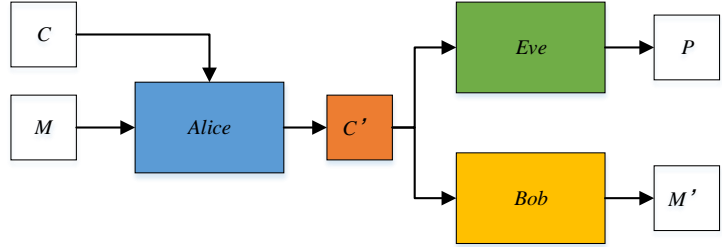

Fig. 15. 3-PLAYERS GAME for steganography by GAN[15].

In Fig.15, Alice uses a cover image, $C$, and a secret message, $M$ to generate a stego image $C'$, Bob try to recover message $M'$ from $C'$. Eve outputs a probability $P$ to indicate the likelihood of a secret message in the image. Alice hopes to learn a steganography scheme in which Eve outputs $P = 1/2$. $A(\theta_A, C, M)$ $B(\theta_b, C')$ and $E(\theta_E, C, C')$ are output for Alice, Bob and Eve, respectively. In order to design a steganographic algorithm by Alice, three loss function $L_A$, $L_B$, $L_C$ are given as the loss of Alice, Bob and Eve.

$$L_B(\theta_A, \theta_B, M, C) = d(M, B(\theta_B, C')) = d\left(M, B(\theta_B, A(\theta_A, C', M))\right) = d(M, M') \quad (12)$$

$$L_E(\theta_A, \theta_E, C, C') = -y\log(E(\theta_E, x)) - (1-y)\log(1 - E(\theta_E, x)) \quad (13)$$

$$L_A(\theta_A, C, M) = \lambda_A d(C, C') + \lambda_B L_B + \lambda_E L_E \quad (14)$$

where $y = 0$ if $x = C'$ and $y = 1$ if $x = C$, $d(C, C')$ is the distance between the C and $C'$, and hyper parameters $\lambda_A, \lambda_B, \lambda_E \in$ R define the weight for each loss term.

Zhang et al.[91] propose an end-to-end model, called STEGANOGAN, for image steganography. They use adversarial training to solve the steganography task and treat the messages embedding and extraction as encoding and decoding problems, respectively. The architecture of STEGANOGAN consists of three sub-modules, as shown in Fig. 16, the image encoder uses the cover image and the message to generate a stego image; a decoder is going to recover the message with stego image, and the quality of the stego image is evaluated by an auxiliary Critic network.

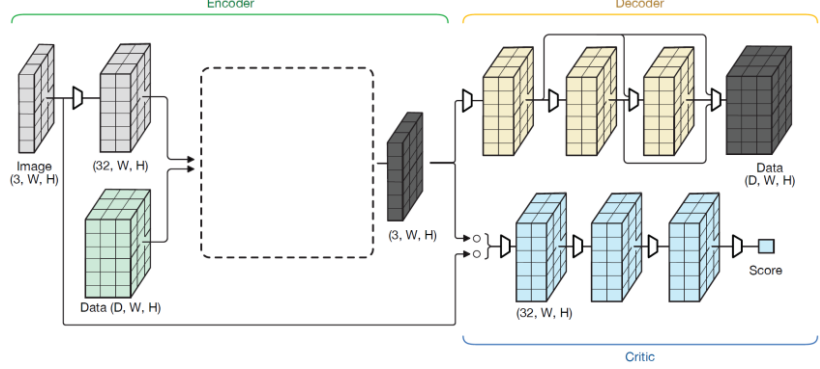

Fig. 16. The architecture for STEGANOGAN model[91.

The training process is divided into two parts. Three losses: the cross-entropy loss $L_d$ for message decoding accuracy, the similarity loss $L_s$ between stego and cover and the realness loss $L_r$ of the stego image using the critic network. The training objective is to

$$\text{minimize } L_d + L_s + L_r. \tag{15}$$

They minimize the Wasserstein loss to train the critic network.

In [92], Zhu et. al also trained encoder and decoder networks to implement message embedding and extraction. They introduces various types of noise between encoding and decoding to increase robustness but focuses only on the set of corruptions that would occur through digital image manipulations. Similar to [92], Tancik et al. [93] achieve robust decoding even under "physical transmission" by adding a set of differential image corruptions between the encoder and decoder that successfully approximate the space of distortions.

Although above algorithms generate stego images through neural networks, it should be emphasized that the ideas of these methods are different from those of cover synthesis steganography introduced in Section 6.3. These algorithms are essentially dependent on a specific cover image, we call it the ***supervised cover synthesis steganography*** (SCSS). The stego image generated by neural network is highly correlated with the original cover, so those algorithms are similar to the cover-modification steganography.

**6.2 Unsupervised cover synthesis steganography**

**6.1.1 Steganography without embedding**

Hu et al. [19] proposed a stego-image synthesis method, steganography without embedding (SWE). In our opinion, a more appropriate term would be "steganography without modification", since embedding operations should be a general term for information hiding operations and should include modification, selection, and synthesis. In their method, the secret messages are mapped into a noise vector is sent to the generator as input to produce a stego image. In this paper, we treat Hu's method [19] as an unsupervised manner which generates stego images only by using the noise itself, we call it the ***unsupervised cover synthesis steganography*** (UCSS). The proposed SWE framework consists of three phases, as illustrated in Fig. 17.

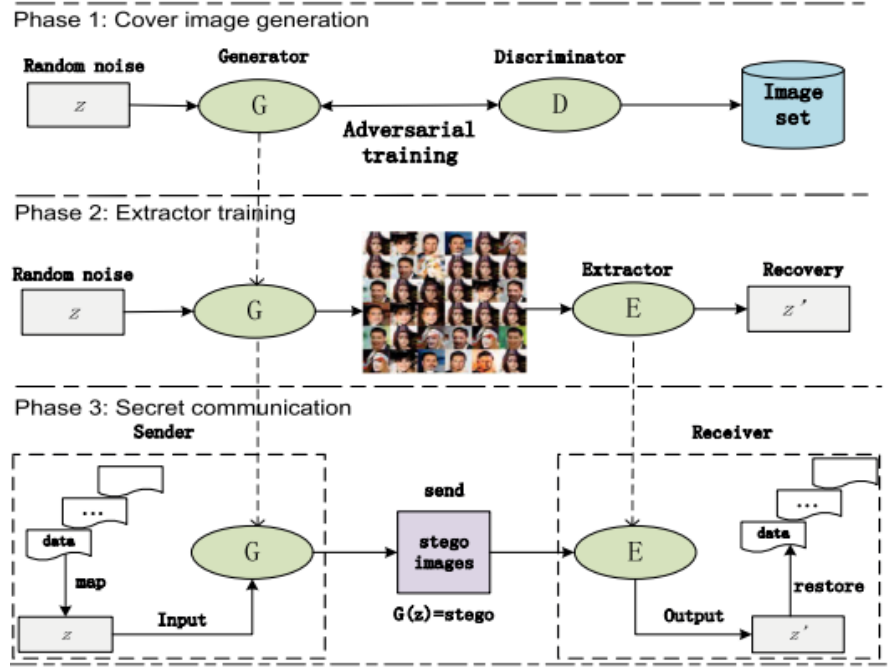

Fig. 17. The framework of SWE method[19]

In a practical scheme, DCGANs is trained to obtain generator G. The generator G is trained with dataset in the first phase, the goal of this phase is to ensure that the generator can produce realistic images., During the second phase, an extractor E is trained with a message extraction loss from a large number of noise vectors, the goal of this phase is to ensure that the noise in the image can be recovered. The loss for extractor training is illustrated as follow:

$$L(E) = \sum_{i=1}^{n}(z - E(\text{stego}))^2 = \sum_{i=1}^{n}(z - E(G(z)))^2 \qquad (16)$$

In the secret communication phase, the sender build a relationship between noise and message, in their scheme, both secrete message *m* and vectors *z* are segmented for mapping. Receiver can uses E to recover noise vector *z* and then the secret message is obtained by the mapping relationship. The highlight of this article is to map the noise to the message so that the message is hidden in the noise. A special extractor is trained to extract noise (message).

**6.1.2 Steganography by WGAN-GP**

Inspired by Hu's method, Li et al. [94] propose a new framework which train the message extractor and stego image generator at the same time. WGAN-GP instead of DCGAN is adapted to generate stego image with higher visual quality. In their method, Generator *G* is trained in a mini-max game to compete against the Discriminator(*D*) and Extractor(*E*) ,as illustrated in Fig.18 . The objective function for training this model is as follows :

$$\min \max \min J(D,G,E) = \{E_{x \sim p_{data}(x)}[D(x)] - E_{z \sim p_z(z)}[DG(z)] + \lambda E_{\hat{x} \sim p_{data}(\hat{x})}[\nabla_{\hat{x}} \|D(\hat{x})\|_2 - 1]\}$$
$$+ \beta\{E_{z \sim p_z(\hat{x})}[\log(z - E(G(z)))]\} \qquad (17)$$

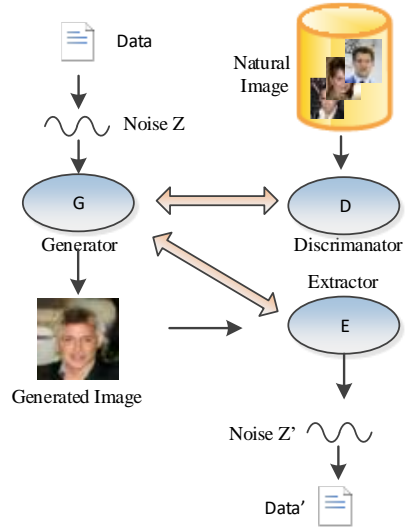

Fig 18. The framework of the steganography with WGAN-GP[94]

where *β* is a positive number which balances the importance of realistic images and correct extraction rate of noise *z*. *λ* is the gradient penalty coefficient.

**6.3 Semi-supervised stego image synthesis**

**6.3.1 Steganography by ACGAN**

In order to allow for semi-supervised learning a steganographic scheme, we can add an additional task-specific auxiliary network in original .Inspired by ACGAN, Liu et al. [95]first proposed a stego-image generation method by ACGAN. This method establishes a mapping relationship between the class labels of the generated images and the secret information, both class label and noise put into the generator for stego image generation directly. We call it the *semi-supervised cover synthesis steganography* (Semi-SCSS)The receiver extracts the secret information from the hidden image through a discriminator.

The ACGAN-based cover synthesis method attempts to establish a correspondence between image categories and secret information. ACGAN for generating the stego image, as illustrated in Fig.19.

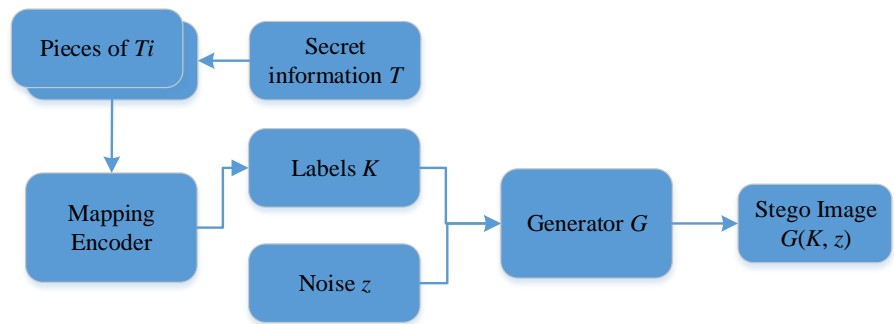

Fig. 19. The framework of steganography with ACGAN

At the message extraction phase, the stego image is fed into a discriminator for getting the pieces of secret information.

Hu's method [19] and Li's [94] methods are essentially the same as Liu's method [95], all of which attempt to create a mapping between the input vector of the generator and the secret message. The former establishes the mapping between noise $z$ and the message, while the latter utilizes the auxiliary control information, such as labels.

**6.3.2 Steganography by Constraint Sampling**

Liu et al.[18, 96] proposed a generative steganography by sampling (GSS). In this scenario, the steganographic embedding operation becomes an image sampling problem. They treated stego image generation as an optimization problem of minimizing the distribution distance between the data image and the cover image:

$$Gen(\boldsymbol{m},\boldsymbol{k}) = \arg\min_{y \sim p_{stego}} D_{JS}(\boldsymbol{p}_{stego}, \boldsymbol{p}_{data}) \tag{18}$$

$$\text{st.} \quad Ext(\boldsymbol{y},\boldsymbol{k}) = C_k \boldsymbol{y}, \tag{19}$$

where Gen(.) is a image generator, and $C_k$ is the secret key $k$. The stego image, $y$, does not depend on any specific cover, which follows the distribution $p_g$, $y = G(z)$.

To implement this solution, they first train an image generator.by DCGAN, as illustrated in Fig. 20(a). The goal of training this generator is to be able to get realistic natural images. Ideally, it reaches an equilibrium state, $p_{stego} = p_{data}$.

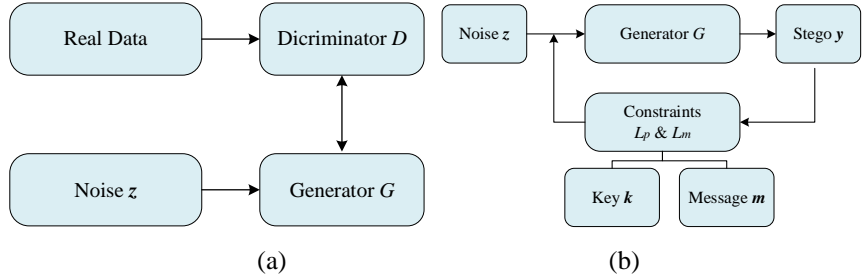

(a)          (b)

Fig. 20. Workflow for GSS (a) Training a image generator; (b) finding a stego image with constraints.

Then, constrained sampling of the image is achieved by defining a message extraction loss constraint. as shown in Fig. 20(b). More specifically, The process of finding a cover image $y$ can be regarded as an optimization problem as follows:

$$\hat{z} = \arg\min_{z}(L_m(z|\boldsymbol{m},\boldsymbol{k}) + \lambda L_p(z)) \tag{20}$$

where $\hat{z}$ is the "closest" encoding of stego image, $L_m$ and $L_p$ denotes the message loss and the prior loss, respectively. Back-propagation to the input noise $z$ is introduced for solving this optimization problem. Under the guidance of this framework, they implemented a digital Carden grille steganography scheme using image completion technology[58].

In this scheme, the image completion technology makes the scheme closer to the idea of Cardan grille, and at the same time, the method becomes a semi-supervised cover synthesis steganography (semi-SCSS) method. In fact, image completion technology is not necessary, that is to say, this scheme can be converted into an unsupervised method, and extended to more image synthesis applications. In this framework, cover synthesis steganography becomes an optimization problem that satisfies both message loss constraint and image perceptual constraint. Unlike Hu's method[19], the GSS framework actually provides an alternative way to cover synthesis steganography using generators. In Hu's article, after the message-noise map and well-trained extractor are ready, the cover image can be obtained by using noise once, while in GSS scheme[96], the stego image carrier is obtained via an iterative sampling method step by step.

**6.3.3 Steganography by Cycle GAN**

In addition to using noise, labels, and corrupted images to generate stego carriers, some researchers treat the cover synthesis as an image-to-image translation problem. Image-to-image translation is a transformation that converts one type of image to another. A very famous model for image translation is is CycleGAN [97]. Although CycleGAN lacks the supervision of the pairing example form, it can take advantage of the supervision at the collection level. CycleGAN is able to convert an image from class *X* through transform *F* to a class *Y*, and can also convert it back to class *X* by transforming *G*. CycleGAN trains transforms F and G by minimizing the adversarial loss $L_{GAN}$ and cycle consistency loss $L_{cyc}$. Chu et al. [98]first claim that CycleGAN can be seen as an encoding process for infromation hiding. By treating CycleGAN's training process as a generator of training adversarial examples and demonstrating that cyclical consistency losses cause CycleGAN to be particularly vulnerable to adversarial attacks

Since CycleGAN's image transformations have some reversible properties, Di et al. [99] proposed a cover synthesis steganography scheme by cycleGAN with reversible properties. Inspired by Hu's framework, they introduced cycleGAN into the new framework and used it for the reversible recovery of the cover image which is generated by a noise vector. Similar to [100], the transformed image can also be regarded as a special encrypted image. In addition, a new extractor is trained to extract the secret data, which also make the data hiding framework reversible. The illustration of Di's method has been shown in Fig. 21.

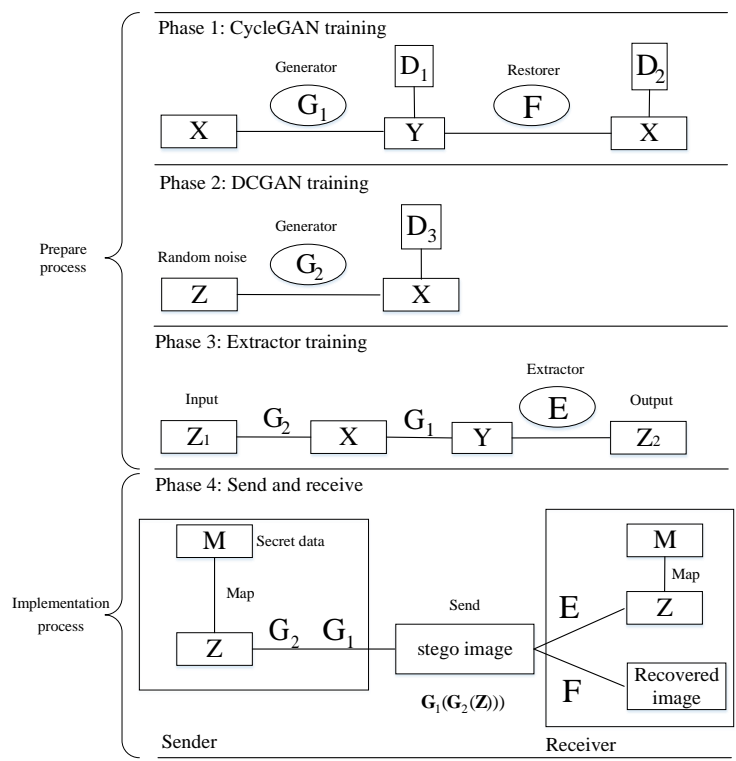

Fig. 21. The workflow of the method [99]

In phase 1, a generator $G_1$ and a restorer F are generated by CycleGan. With two discriminators $D_1$ and

$D_2$, two transformations achieved $G_1$: X->Y and F: Y->X, where X and Y are image collections. In Phase 2, a generator $G_2$ is generated by DCGAN method with the help of discriminator $D_3$. In Phase 3, based on the two discriminators $G_1$ and $G_2$ we can get the transformation from random noise to stego image set Y. Then, a new extractor E is trained with a neural network which ensures that the generated output $Z_2$ is same as the input $Z_1$ as closely as possible. Before data hiding, the sender sends extractor E and restorer F to the receiver. Both sides learn a mapping from secret data *M* to noise Z. Corresponding to the traditional RDH methods, the image generated by G1 and G2 can be regarded as a cover image and marked image. Then, the sender sends the marked image $G_1G_2(Z)$ to the receiver. At the receiver side, recover image can be obtained and the embedded data can be extracted.

## 6.4 Summary on Stego Image Synthesis

Currently, although the works of literature on generating stego images with generators is not a lot, they are very attractive and representative. In this section, we will further analyze the characteristics of these methods and summarize some general rules. In this article, we also refer to image synthesis steganography as *generative steganography*, which refers to the methods of directly obtaining a stego image by a generator without a specific original cover image.

### 6.4.1 Sender mode

With GAN's generator, realistic images are sampled from the distribution of a dataset. This avoids the problem of establishing an explicit distribution model for real images. Sampling a stego image from a generator makes the steganography problem a sampling process. According to the basic principles of steganography implementation strategy described in section 2.3, cover synthesis steganography also has the following two implementation strategies.

**Payload-limited Sender**: In practice, it is difficult to achieve the optimal which satisfies $p_g=p_{data}$. In the case that the message length is limited to *m* bits, the cover synthesis can be regarded to minimize the distance between $p_{stego}$ and $p_{data}$:

$$\text{Emb}(m,k) = \underset{y \sim p_{stego}}{\arg\min} D(p_{stego}, p_{data}) \tag{21}$$

$$\text{st. } \text{Ext}(\text{Emb}(m,k),k) = m, \forall m \in \{0,1\}^m \tag{22}$$

**Distance-limited Sender**: Due to the randomness of the images generated by the generator, when the distance between the distribution of the cover image and the real data distribution is within an acceptable range, the cover synthesis steganography can also be regarded as an optimization problem to maximize the capacity of the message:

$$\text{Emb}(m,k) = \underset{y \sim p_{stego}}{\arg\max} |\text{Ext}(\text{Emb}(m,k),k)|_m \tag{23}$$

$$\text{st. } D(p_{stego}, p_{data}) < \varepsilon \tag{24}$$

where $|\cdot|_m$ denotes the length of message *m*. This optimization problem indicates that the goal of cover synthesis steganography in this mode is to increase the capacity of the generated image under the premise of satisfying some indistinguishable metric.

This framework has three differences compared to the framework which minimizes distortion. First,

this scheme directly minimizes the distribution distance rather than the distortion caused by modifying operation. Second, the scheme is very easy to introduce a secret key, making the scheme meet the Kerckhoffs' principle. Third, the traditional minimum distortion algorithm usually adopts Payload-limited sender method to design the steganographic scheme. Actually, the more intuitive use of steganography should be Distance-limited Sender mode. Although formally, this mode is very similar to the mode with distortion-limited, but there is a fundamental difference between the. The relationship between distortion and steganography security is ambiguous. The process of training generative model is theoretically reducing the distribution distance, which makes the distance-limited mode theoretically supported. In fact, all of these methods, including ACGAN-based method[95], SWE method[19, 94], cycleGAN-based method[99] and GSS method[96], introduce message-noise mapping or message loss constraint after the generator is trained. The trained generator represents that a fixed distribution distance, and the message mapping or message loss constraint aim to improve the embedding capacity, all these methods adopt the distance-limited sender mode.

**6.4.2 Message embedding and extraction**

The goal of cover synthesis steganography is to generate realistic images while hiding messages. Traditional GANs focus on finishing the realistic image generation task. For steganographer, the most critical task is how to embed and extract messages. Interestingly, we will see that, in contrast to traditional steganography schemes which focus on the design of embedding operation including modification and selection. In the cover synthesis steganography based on the generator, message extraction and embedding operations will be treated as a single issue. In some circumstances, we will pay more attention to the extraction strategy of messages. The task of information hiding becomes the challenge of whether the message can be extracted correctly.

Under the framework of cover synthesis, image steganography becomes the task of the space mapping between message space M and stego image space S. The embedding process is regarded as a message-stego mapping, while the message extraction can be regarded as a stego-message mapping, as shown in Fig. 22. Because of the randomness of generating secret carrier based on the generator, the mapping relationship between message and secret carrier may be one-to-many, and the goal of steganography is to seek the mapping relationship satisfying the constraints of message loss and realness loss.

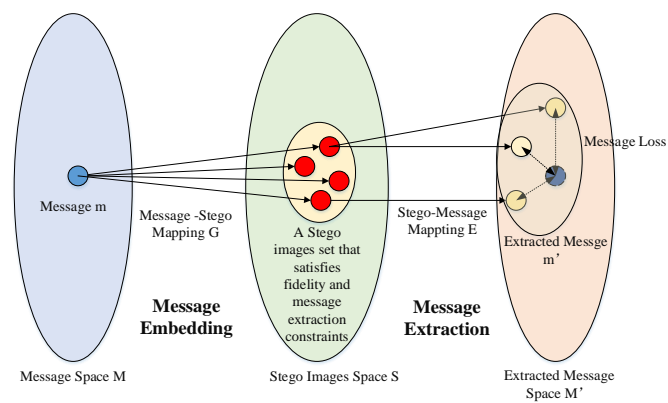

Fig. 22. A framework for messages embedding and extraction in cover synthesis by GAN

In fact, similar to the spatial domain and transform domain steganography, the locations of messages in the stego image is different. For generative steganography, when the message loss constraint directly acts on the space domain of images, this type of steganography is a spatial domain generative steganography, such as the Liu's method [96], and the secret information is directly hidden in the generated pixel itself. When the message constraint acts on the transform domain, cover synthesis steganography can also be regarded as a transform domain steganography scheme. In Hu's method [19], they hide the message in noise and needs to be recovered by a neural network extractor. Similar to [19], [95] hides the message in the semantic labels. These methods are all steganography schemes in the transform domain. In addition, the deep learning model is regarded as an encoder, which converts data into a feature space, and a method of steganography in the encoding process, such as [93], is also a kind of steganography in the transform domain. One of the advantages of transform domain steganography is that the encoded messages can resistance image distortions of various forms. Although robustness is not the goal of traditional steganography, robust steganography is seen as a practical requirement in some specific situations.

Therefore, in the case of using a neural network or generator, the steganography is converted into an optimization problem of defining a total loss function,

$$L_{\text{total}} = \lambda_p L_{\text{perceptual}} + \lambda_m L_{\text{message}} + \lambda_{ro} L_{\text{robustness}} + \lambda_{re} L_{\text{reversible}} \qquad (25)$$

where $L_{perceptual}$ and $L_{message}$ represent the concerns of traditional steganography: the accurate extraction of the message and the natural properties of the stego image. The latter two losses $L_{robustness}$ and $L_{reversible}$ represent some other properties such as robustness or reversibility in steganography. These Loss weights λs indicate the proportion of each performance requirement in different application scenarios.

**6.4.3 Strategy: from supervised to unsupervised**

According to the different information received by the generator when generating the stego image, we divide cover synthesis method into supervised, unsupervised, semi-supervised in Section 6.1-6.3 respectively. We have found that as a semi-supervised method, the method of constrained sampling [96] using corrupted images can be regarded as a general framework for stego image generation. Both supervised methods and unsupervised methods can be seen as a special case of constrained sampling methods.

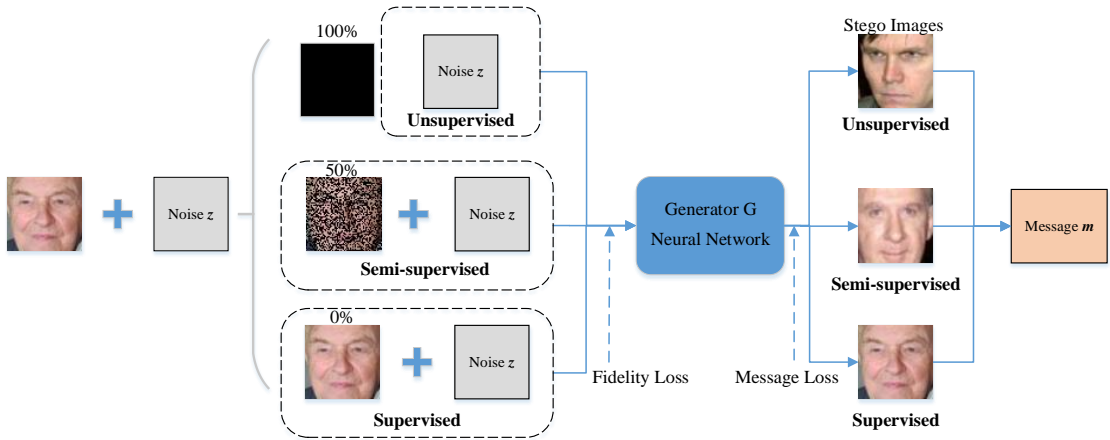

Fig. 23. A general framework for stego image generation

Under this framework, we can relatively easily grasp the commonalities of these three strategies. In unsupervised method[19], there is no original cover image that can be treated as 100% corruption.

Therefore, it is necessary to construct a stego image by utilizing the mapping relationship between message and noise, the message loss constraint is based on the noise extraction accuracy. In semi-supervised method[96],with image completion techniques, secret messages are embedded in uncorrupted image regions, and message loss constraints are built into a portion of the image. In supervised method[15, 78], due to the existence of the original carrier, the constraints of the message loss are based on the difference between the generated stego images and the original images.

The mechanism of obtaining the synthetic stego image dependents on the training mode of the generative model, which can be divided into two implementation strategies. The first strategy is to train a generator with message loss constraint and prior constraint simultaneously, namely parallel constraint synthesis (PCS) mode, as shown in Fig. 24. After the generator is trained, the cover image can be sampled directly from the generator. However, because the message is relevant to the generator. You need to repeatedly train a new generator when a new message needs to be hidden. Currently, due to the high cost of training generators, this strategy has great limitations in practice. To the best of our knowledge, Li's method[94] follows this framework that utilizes parallel constraint synthesis mode.

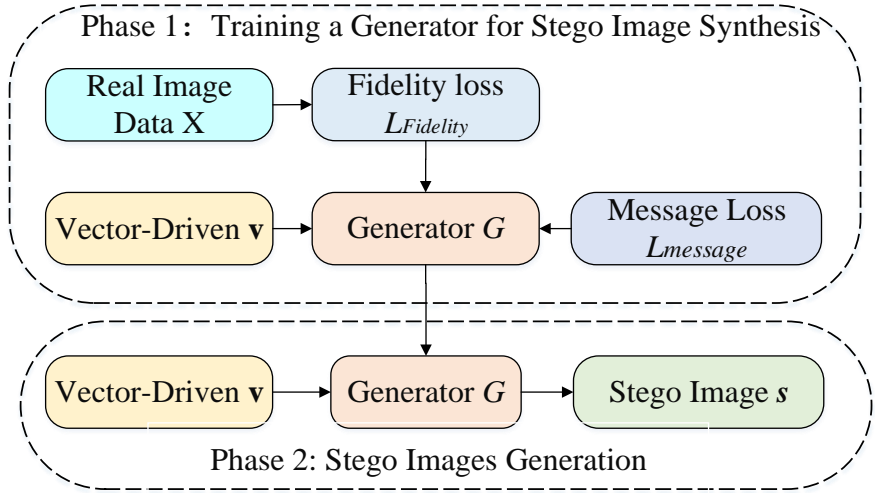

Fig. 24. Constraint parallel training mode for the cover synthesis

The second strategy is to satisfy the prior constraint and message constraint respectively through two sequential schemes, namely sequential constraint synthesis (SCS) as shown in Fig. 25. First, a real image data set is used to train a generator that satisfies the realness loss constraint, $L_{Fidelity}(s)$. Then, we can design a generation scheme that meets the message extraction loss constraint such as $L_{mesage}(m_{ext}|s,k)$. The character of this scheme is that the constraints of message loss and prior loss can be separated, that is, when training the generator, we only need to pay attention to how to make the generated sample distribution approximate to the real data distribution. After training, the generator is used to construct the candidate stego image and the final secret carrier is obtained by using message constraint. The separability of constraint conditions will make the design of cover synthesis steganography more simple and practical. This separation is actually a specific implementation scheme of the payload-distance sender mode. All of these methods such as [19, 95, 96] above all adopt this serial mode.

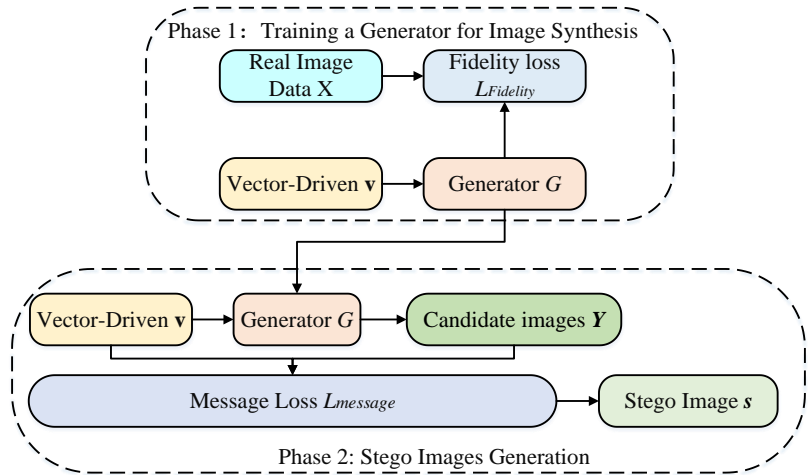

Fig.24. Separate serial constraint mode for the cover synthesis

## 7 Evaluations Metrics on GAN-based Steganography

In this section, we evaluate GAN-based steganography on three axes: *secrecy*, the difficulty of detecting stego images; *capacity*, the number of message bits that can be hidden in the stego image.; and *robustness*, the degree to which methods can succeed with some image distortions. All mentioned methods are divided into three types such as cover modification, selection, and synthesis as before.

### 7.1 Security

Steganographic security mainly includes indistinguishability and computational complexity of intercepting secret keys. In this section, we start with image quality evaluation.Then, we compare the statistical indistinguishability of stego images obtained by these methods via data-driven steganalysis tools.

#### 7.1.1 Image quality

First of all, it should be noted that the cover selection steganography [16] only selects an original cover image as a stego image, we assume that the image quality of this method is perfect. Our comparison is mainly directed to the methods of constructing stego images by a generator, which includes cover modification and synthesis. We report the experimental results from qualitative and quantitative comparisons.

*Qualitative comparisons* In Table 1, we present the generators used in the different cover modification methods and the image datasets, as well as the visual effects of the stego images obtained by these methods. As can be seen from the table, those steganographic methods, such as [17, 79, 81], that use the generator to generate the original cover image, the resulting stego image quality is not good. This is mainly due to the stego image quality depending on the performance of the generator. And those methods such as[14, 82], that use the GAN to learn a modification probability matrix have higher visual effects because they rely on the original carrier.

Table1. Comparisons of images for cover modification by GANs

| Ref. | Generator | Dataset | VISUAL Quality |
|---|---|---|---|
| [79] | GAN | CelebA[101] | 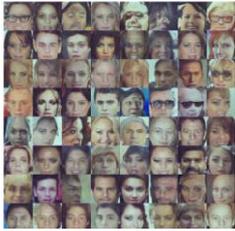 |
| [17] | WGAN | CelebA[101] | 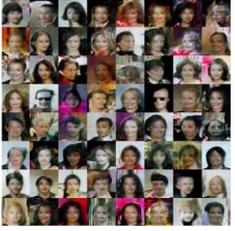 |
| [81] | WGAN | CelebA[101] | 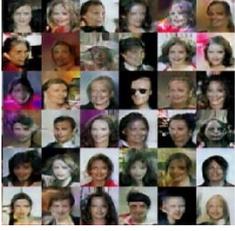 |
| [15] | Alice Encoder: four layers are a sequence of fractionally-strided convolutions, batch normalization and ReLU, except for the final layer where tanh is used as the activation function. | BOSS[102] and CelebA[101] | 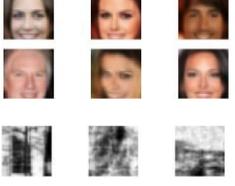 |
| [82] | ASDL-GAN | BOSS[102] | 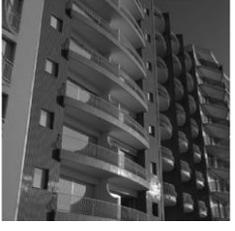 |
| [14] | UT-SCA-GAN | BOSS[102] | 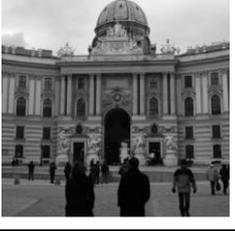 |
| [91] | Critic network | Div2k[103] and COCO[104] | 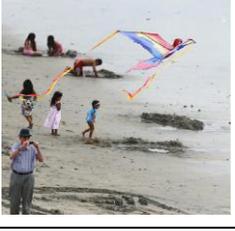 |

Table 2 shows the synthesis methods for directly generating a stego image using a generator without relying on a specific original cover image. From these methods, it can be seen that the visual quality of the stego image is completely dependent on the performance of the generator. The generators used by these methods are relatively simple, so the resulting effects are not good. The one exception is that in CycleGAN-based steganography [99], BEGAN is used to generate images, so it has a higher visual quality.

Table2. Comparisons of images for cover modification by GANs

| Ref. | GAN model | Dataset | VISUAL Quality |
| --- | --- | --- | --- |
| [95] | ACGAN | MNIST[105] and CelebA[101] | 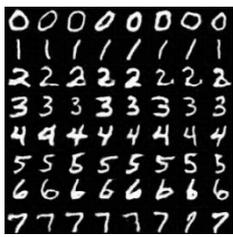 |
| [19] | DCGAN | CelebA[101] and Food101 [106] | 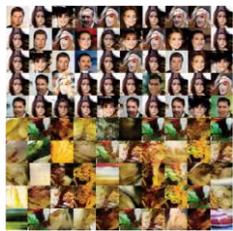 |
| [96] | DCGAN: | LFW[107] and CelebA [101] | 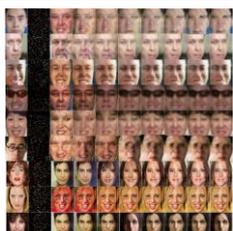 |
| [99] | CycleGAN: | horse2zebra, woman2man [99] | 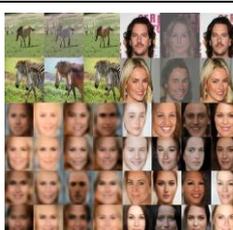 |
| [98] | CycleGAN | 1,000 aerial photographs X and 1,000 maps Y[98] | 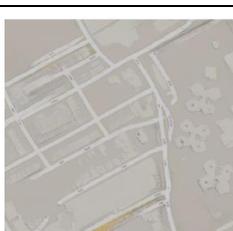 |

*Quantitative analysis* One widely-used metric for measuring quality of image is the peak signal-to-noise ratio (PSNR) and Structural similarity index (SSIM) [108] between the cover image and the sego image. Since GAN-based cover modification uses LSB-like [109, 110] or minimizes distortion[32] in the modification strategy, it has been shown that these methods make the PSNR value large, and the image quality difference is small compared with the original image. The *SteganoGAN* method[91] reports the PSNR and SSIM in their work, where the PSNR values fall between 35-41 and the SSIM values are above 0.9. However, for a method of directly generating a dense image using a generator, there is no one-to-one pixel correspondence. Metrics like PSNR are not suitable to evaluate the stego image. Quantitative indicators for GAN model often use Fréchet inception distance (FID) [111] and inception score (IS)[5]. Other evaluation criteria include Mode Score [56], Kernel MMD [112], Wasserstein distance, and 1-nearest neighbor (1-NN)-based two sample test [113]. These indicators are still an ongoing important research area.

**7.1.2 Statistical Steganalysis**

Steganographic security is often evaluated using a steganalyzer to distinguish between cover and stego images. In this paper, we directly adopted the best results reported in their original article. Since these methods use different ways for steganalysis, we also point out the classifiers they use while giving the detection rate. In this case, although we cannot evaluate the performance of these algorithms objectively, their experimental results will give us a relative criterion for the security of these methods.

Table 3. The FIDs of different models trained on CelebA

| Methods | Embedding methods | Classifier | Error rate |
| --- | --- | --- | --- |
| SGAN [79] | ±1 embedding | *Self-Defined* | 0.50 |
| SGAN[79] | HUGO | *Self-Defined* | 0.49 |
| SSGAN [17] | ±1 embedding | *Self-Defined* | 0.72 |
| SSGAN [17] | HUGO | Qian's Net [80] | 0.71 |
| ASDL-GAN[82] | minimal-distortion | Xu's Net [84] | 0.27 |
| ASDL-GAN[82] | minimal-distortion | EC+SRM[114, 115] | 0.26 |
| UT-SCA-GAN[14] | minimal-distortion | EC+SRM[114, 115] | 0.22. |
| UT-SCA-GAN[14] | minimal-distortion | EC+maxSRMd2[114, 115] | 0.21 |
| Tang[88] | Adversarial Embedding | Xu's Net [84] | 0.58 |

It can be seen from Table 3, those methods that use the generator to generate the original cover and introduce the steganalyzer have good security for their own steganalysis tools, such as [79] [17]. At the same time, the security of using GAN to construct a modification probability matrix is very close to the traditional cover modification steganography algorithm, such as [82] [14].

Table 4. The FIDs of different models trained on CelebA

| Methods | Embedding methods | Classifier | Error rate |
| --- | --- | --- | --- |
| Hayes[15] | Adversarial Trainning | *Self-Defined* | 0.79 |
| Zhang[91] | Adversarial Trainning | Ye et al.[116] | 0.50 |
| SWE[19] case 1 | Message-noise | Ni's model [116] | 0.53 |
| SWE[19] case 2 | Message-noise | Ni's model[116] | 0.02 |
| ACGAN[95] | Message-label | EC+SPAM[114, 117] | 0.52 |
| GSS[96] | Message Loss | EC+SPAM[114, 117] | 0.42 |
| GSS[96] | Message Loss | EC+SCRMQ1[114, 118] | 0.04 |
| CycleGAN [99] | Message-noise | Ni's model [116] | 0.54 |

In Table 4, we list the error rate of statistical analysis of GAN-based cover synthesis steganography. Based on the original cover image as an input, those methods that use the adversarial game strategy to generate stego images, such as [15] [91], also have a certain security. SWE's[19] case 1 assumes that the attacker is unable to obtain training samples, and the security is higher at this time, but in case 2, when directly using training images to train the steganalyzer, steganalysis achieve good detection ability. The problem with this approach is that the steganographic analysis becomes forensic of the composite image at this time, that is, whether the image is a synthetic image. ACGAN-based method[95]only considers cases where training samples cannot be obtained. In the GSS[96]  method, under the embedded capacity of 0.4bpp, the security is higher for SPAM features, but for SCRMQ1[118], classifies achieve good detection ability. The benefit of this sampling method is that the training set can be exposed, and the steganographic security can depend on the confidentiality of the embedded key.

**7.1.3 Security levels with Kerckhoffs's principle**

Ke et al[28]. proposed a stego-security classification strategy with Kerchhoffs's principle based on the different levels of steganalysis attacks such as Stego-Cover Only Attack (SCOA), Known Cover Attack (KCA), Chosen Cover Attack (CCA) and Adaptive Chosen Cover Attack. In synthesis methods, such as [19, 95], there are explicitly extraction or embedding key $k$. In fact, the mapping itself can be used as a key, but in this case, the key space is too small to resist SOA attacks. Therefore, when the algorithm exposes an active attack environment that directly attempts to extract a key, it is not secure in terms of the computational complexity of acquiring keys. In GSS method [96], the keyspace meets the certain computational complexity when the size of Cadan grille are large enough. Therefore, the GSS method can be stego-secure against SCOA. The training image set should be available for the attacker in KCA model. In [19], it has been shown that directly using the training set to train classifies for steganalysis is unsafe. Therefore, cover synthesis method is not stego-secure against KCA. At present, the actual security requirements for cover synthesis are as follows.1) the training dataset and the key $k$ should be kept secrecy. 2) |$\mathbb{K}$| should be large enough to meet certain requirements of computational complexity. .

**7.2 Capacity and Recovery Accuracy**

At present, in terms of steganographic capacity, there is still a big gap between cover synthesis steganography and traditional cover modification steganography. It has reached a considerable level compared to the traditional cover selection or synthesis method. We list the capacity of cover synthesis methods by GAN with other selection or synthesis methods in Table 5, the absolute capacity is shown in the second column, the size of the stego image is listed in the third column, the relative capacity is listed in last column:

$$Relative\ capacity = Number\ of\ message\ bits/Size\ of\ the\ image \qquad (26)$$

Table 5. Capacities of various non-modification methods.

| Reference | Absolute Capacity (bytes/image) | Image Size | Relative Capacity (bytes/pixel) |
|---|---|---|---|
| [119] | 3.72 | ≥512×512 | 1.42e-5 |
| [43] | 1.125 | 512×512 | 4,29e-6 |
| [120] | 2.25 | 512×512 | 8.58e-6 |
| [51] | 64×64 | 800×800 | 6.40e-3 |
| [49] | 1535~4300 | 1024×1024 | 1.46e-3~4.10e-3 |
| [19] | ≥37.5 | 64×64 | 9.16e-3 |
| [95] | 0.375 | 32×32 | 3.7e-4 |
| [96] | 18.3-135.4 | 64×64 | 1.49e-3~1.10e-2 |

*Extraction rate* However, in those schemes, such as [15, 19, 94-96], where the stego image is directly generated by the generator, the stego image generation depends on the optimization problem of the neural network, that is, the minimization of a certain cost function. Since the generator or neural network usually cannot get the optimal solution, the message may not be extracted correctly, the actual capacity should be calculated as:

$$Actual\ capacity = Relative\ capacity \times Extraction\ rate \quad (27)$$

In Fig.26 we show the extraction rate of different algorithms when the message cannot be completely extracted.

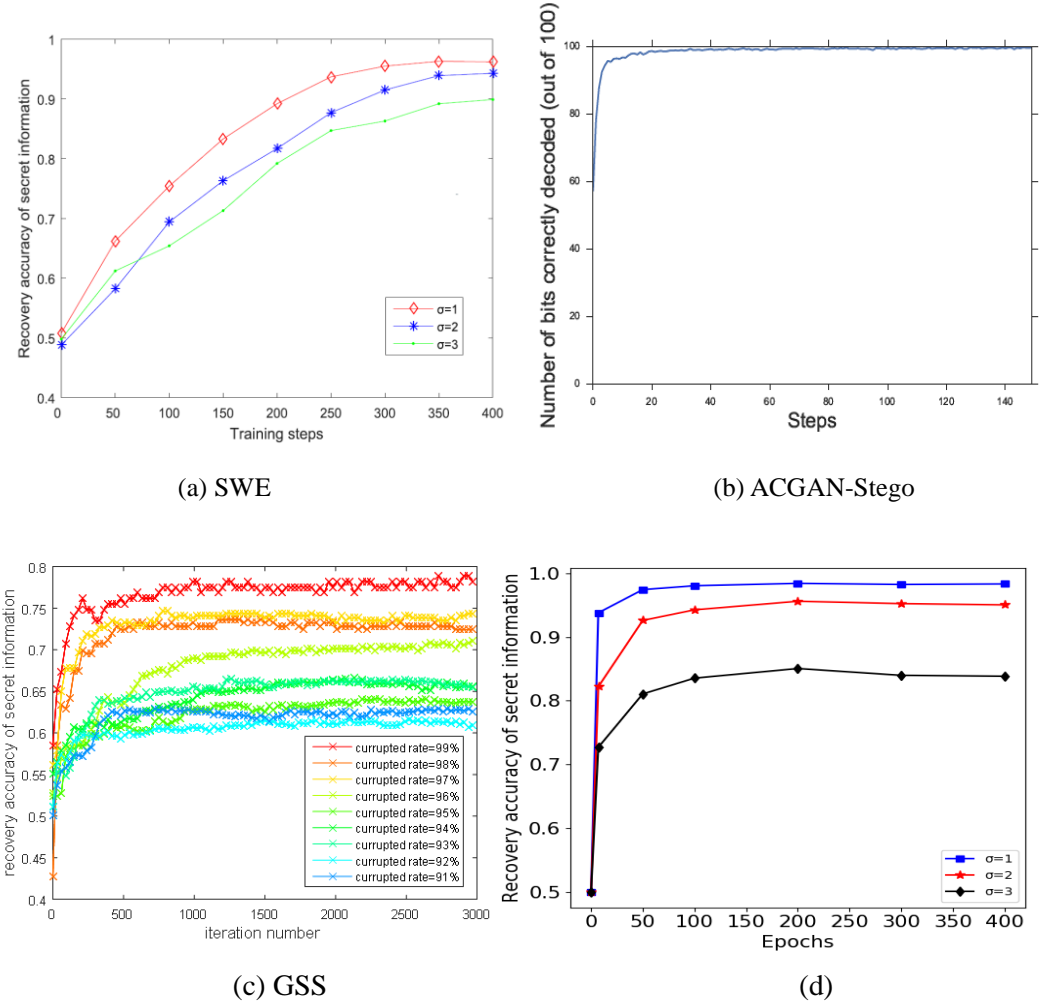

(a) SWE  (b) ACGAN-Stego

(c) GSS  (d)

Fig. 26 Extraction rate for cover synthesis steganography

It can be seen from the Fig.26 that in the methods of (a)[19] and (b)[95], additional training is required, and the message extractor gradually increases the stability of the message extraction as the number of training steps increases. The ACGAN-based method message is hidden in the label of the image, that is, the semantic level. By training a classifier, the category information can be obtained, so that the message extraction accuracy is high, and the disadvantage is that the sneaking rate is low. In Fig.26 (c) [96], in order to verify the accuracy of message extraction, we first perform random damage, according to the damage ratio of 99%-91%, and embed the message on all pixels that are not damaged. The recovery accuracy increases as the iteration numbers increases. Due to limitations in learning performance that the message constraint cannot be completely satisfied. The actual embedding capacity is not high. In Fig.26 (d), the recovery accuracy[94] increases as the number of training steps increases. After about 10 epochs (every epoch has 633 steps), the recovery accuracy rapidly increased to a higher level.

**7.3 Robustness**

In the cover synthesis steganography such as [19] and[95, 121], the message is actually hidden in the transform domain of the generated image, so that it has certain robustness. Furthermore, by varying the types of image distortion during training process, [92] and [93] show that steganography model can learn robustness to a variety of different image distortions. In this section, we only test the robustness of [19] and [121] to common image attacks. We consider applying four typical image attacks. These attack conditions are listed as follows.

C1. Contrast enhancement by multiplying the intensity of the image pixels with factors of 1.1 and 1.5
C2. Gaussian noise addition (variance 0.01).
C3. Salt noise added (density 0.05).
C4. JPEG compression with varying quality facts (q.f. 90, q.f. 60 and q.f. 30).

We use the G network to generate 5,000 stego images based on the CIFA-100 dataset and apply the four typical methods to attack each group of images. Then, we give the accuracy of the message extraction after the attack. For Hu et al.[19], the result is shown when parameter σ is 3 and δ is 0.001. A group of results is shown in Table 6.

Table 6. The extraction accuracy of extractor for the attacked stego images

| Attack Condition | C1 | | C2 | C3 | C4 | | |
|---|---|---|---|---|---|---|---|
| | 1.1 times | 1.5 times | | | q.f.50 | q.f.70 | q.f.90 |
| [121] accuracy(%) | 100 | 98 | 98.99 | 99.96 | 98.52 | 99.89 | 100 |
| [19] accuracy(%) | 81.72 | 71.23 | 54.49 | 53.68 | 61.33 | 62.82 | 66.10 |

From the experimental results, these methods are robust to all four attacks, especially the method of Zhang et al.[121], able to resist jpeg compression and contrast enhancement. The model has no errors at a jpeg compression factor of 90 and brightness changes at an intensity of 1.1 times because the message relies on the recognition of image semantic labels by neural networks. The neural networks have good fault tolerance. When inputting fuzzy or incomplete information, a suboptimal approximate solution can be given to achieve correct identification of incomplete input information. The noise extracted from the image has no clear semantic meaning. However, it can be seen from the experimental results that the noise can be resistant to contrast enhancement, but is less robust to JPEG compression. This is consistent with our idea of treating this method as a kind of information hiding in the transform domain.

# 8 Prospective and Conclusion

## 8.1 Perspective

This paper reviews the recent research on image steganography based on GAN. At present, the modified method has obvious advantages in terms of embedded capacity, anti-statistical analysis and message capacity. There is still a long way to go before generator-based steganography can achieve the current steganographic performance. We believe that the methods based on stego image synthesis by GAN will be a promising field of research in steganography. In this section, we briefly discuss the impact of this new steganography on traditional steganography and its development, mainly in the following aspects:

*Steganography in computer vision.* Traditional image steganography is a special method of image processing, rarely attracts attention in the field of computer vision. With the development of generation models, Image steganography begins as a special computer vision application, such as image synthesis or image translation. Based on the inherent consistency of the cover synthesis steganography and generative models, image steganography has become an important application of computer vision. Traditional computer vision researchers have also begun to study image steganography [92]. Combination of the research fields broadens the application areas of steganography. Especially in the future, using a computer to synthesize images, video or other media will be a normal state, and hiding messages in the generated medium will become a new type of covert communication means. In this case, drawing on more advanced research results based on artificial intelligence and computer vision will make the generated steganographic images more realistic and safe. In addition, the introduction of carrier synthesis into the research ideas of information hiding will also have an important impact on the development of other information hiding technologies, such as digital watermarking technology.

*Steganography Capacity* Currently, in generative steganography, such as [96], since the message extraction is performed directly on the image spatial domain pixel values, the uncontrollability of generated image pixels results in message loss, so that the message cannot be effectively extracted. The message in method [19, 95] does not exist on the pixel value itself but exists as a category attribute[95] or a noise vector [19]. Therefore, the message has certain robustness, and the disadvantage is that the embedded capacity is low. In the generative steganography, further improvement of message stability or embedding capacity will be the focus of future research.

*Image evaluation* It is very hard to quantify the quality of synthetic images, in the field of image synthesis, the evaluation criteria of the generated images are not sound enough. Some methods using manual evaluation are subjective and lack objective evaluation criteria. The current evaluation criteria are mainly IS (Inception score) and Frechet Inception Distance (FID). These methods only consider the authenticity and quality of the image. These indicators are still an ongoing important research area.

*Steganalysis* Under the generative steganography framework, the task of steganalysis is divided into two phases. The first phase is image forensics, which distinguishes whether it is a generated image. The second stage is image steganalysis, which distinguishes whether the generated image contains a secret message. The current generated image is only effective in visuals. To distinguish between natural images and generated images, many image forensics methods can be used.

**8.2 Conclusion**

In this paper, we reviewed image steganography with GAN. We first gives the principle and characteristics of the steganography, reviews the traditional image steganography method and the problems faced, and then introduces the principle and some improvement models of GAN. This paper focuses on the GAN-based steganography methods, which are classified according to the framework of steganography based on cover modification, cover selection and cover synthesis. The GAN-based modification method uses GAN to construct the original carrier or a modification matrix. The GAN-based carrier selection method has a low embedding capacity and requires a secret channel to pass the key, thereby reducing its actual availability. The GAN-based cover synthesis method directly uses the GAN training generator to obtain the stego images. We divide cover synthesis by generator into three categories based on the dependence of the cover image, unsupervised, semi-supervised and supervised methods. From the above introduction, we can see that the GAN-based cover synthesis steganography can make full use of the powerful generator, and provide a new implementation method for cover synthesis steganography.